\documentclass[aps,prl,twocolumn,groupedaddress]{revtex4}
\usepackage[dvips]{graphics, color}

\usepackage{graphicx}
\usepackage{dcolumn} 
\usepackage{bm}

\begin{document}

\title{Probing Pair-Correlated Fermionic Atoms through Correlations in Atom Shot Noise}

\author{M. Greiner}
\email[Email: ]{markus.greiner@colorado.edu}
\homepage[URL:]{http://jilawww.colorado.edu/~jin/}
\author{C. A. Regal}
\author{J. T. Stewart}
\author{D. S. Jin}
\thanks{Quantum Physics Division, National Institute of Standards and Technology.}
\affiliation{JILA, National Institute of Standards and Technology
and University of Colorado, and Department of Physics, University
of Colorado, Boulder, CO 80309-0440}

\date{\today}

\begin{abstract}
Pair-correlated fermionic atoms are created through dissociation of
weakly bound molecules near a magnetic-field Feshbach resonance. We
show that correlations between atoms in different spin states can
be detected using the atom shot noise in absorption images.
Furthermore, using time-of-flight imaging we have observed atom
pair correlations in momentum space.
 \end{abstract}


\maketitle

A variety of fascinating quantum systems have been realized with
ultracold atoms over the past decade. A unique feature of these
systems is that the quantum state is very accessible. Density
distributions and matter waves can be directly probed in
time-of-flight (TOF) expansion with absorption imaging
\cite{Anderson1995a}. This method has, for example, allowed
observation of Bose-Einstein condensates (BEC), matter-wave
interference patterns, and quantized vortices in the macroscopic
matter wave. However, experiments are now beginning to access a
further class of quantum systems that involve quantum entanglement
and correlations. Examples include the Mott insulator state for
atoms in an optical lattice \cite{Jaksch1999a,Greiner2002a},
proposed quantum Hall-like states for rapidly rotating condensates
\cite{Cooper2001a}, and condensates of generalized Cooper pairs
\cite{Regal2004a,Zwierlein2004a,Chin2004a}.

Altman \textit{et al.} \cite{Altman2004a} recently pointed out that
atom cloud absorption images can hold information beyond the
first-order correlation provided by the density distribution. They
proposed that density-density correlations can be directly measured
by carefully analyzing the atom shot noise present in TOF
absorption images of the atom gas. This analysis can reveal key
properties of strongly correlated states of atoms such as fermionic
superfluids or exotic states in optical lattices. The method of
detecting two-particle correlations is reminiscent of the
measurement of temporal current noise in mesoscopic conductors
\cite{Blanter2000a} and closely related to the detection of
entangled photon pairs in quantum optics \cite{Scully1997a}, which
is fundamental to experiments studying nonclassical states of
light.

Here we report on the creation and detection of local and nonlocal
pairs of fermionic $^{40}$K atoms. The pair-correlated atoms are
created by dissociating weakly bound diatomic molecules near a
Feshbach resonance and detected through the measurement of atom
shot noise correlations in TOF absorption images. This novel
detection method provides a new tool for probing highly correlated
quantum states in atomic gases. In addition, pair-correlated atoms,
such as demonstrated here, have uses in quantum information,
precision measurements \cite{Orzel2001a}, and fundamental tests of
quantum mechanics \cite{fry1995}.

\begin{figure}[h]
\begin{center}
\includegraphics[width=1\linewidth]{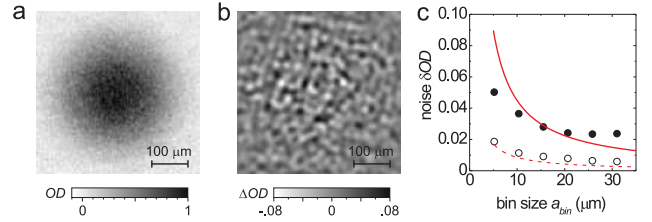}
\caption{Atom shot noise in a time-of-flight (TOF) absorption
image. (a) One spin state of a weakly interacting, two-component,
degenerate Fermi gas with 2.3$\times10^5$ atoms per spin state is
imaged after 19.2\,ms of expansion. (b) The noise on the absorption
image was extracted using a filter with an effective bin size of
15.5 microns. (c) The noise at the cloud center ($\bullet$) is
dominated by atom shot noise, while the noise at the edge of the
image ($\circ$) shows the photon shot noise. The noise in $OD$
decreases when averaged over a larger bin size. The predicted
dependence for atom and photon shot noise is shown by the solid and
dashed lines, respectively.}
\end{center}
\end{figure}

Measuring atom-atom correlations via the method proposed by Altman
\textit{et al.} \cite{Altman2004a} requires that atom shot noise
dominates over other noise sources in absorption imaging. Atom shot
noise arises because of the quantized nature of the atoms and
causes a granularity in the observed density distribution
(Fig.\,1). In absorption imaging, the atoms scatter light out of a
resonant laser beam, and the resulting shadow is imaged on a
charge-coupled-device (CCD) camera \cite{Anderson1995a}. The
optical density is determined using
$OD(\bm{r})=\log(I_\mathrm{ref}(\bm{r})/I(\bm{r}))$, where
$I(\bm{r})$ is the spatial intensity of the shadow image and
$I_\mathrm{ref}(\bm{r})$ is a reference image taken with no atoms.
The atom column density is given by $n(\bm{r})=OD(\bm{r})/\sigma$,
where $\sigma=3\lambda^2/(2\pi)$ is the absorption cross section
and $\lambda$ is the wavelength of the imaging light. There can be
three types of noise in these shadow images: atom shot noise,
photon shot noise, and technical noise. Having a large photon
number per pixel, $N_p$, in the absorption imaging beam keeps the
technical noise of the CCD camera (dark current and readout noise)
significantly smaller than the photon shot noise, $\delta
N_p\!=\!\sqrt{N_p}$. Further, if each atom scatters a large number
of photons, then the atom shot noise, $\delta N_a\!=\!\sqrt{N_a}$,
will dominate over the photon shot noise \footnote{It is important
to minimize technical noise from interference patterns in the
imaging laser beam. Interference is canceled by normalization to a
reference image with no atoms. To optimize the normalization, we
use a relatively clean imaging beam, take the normalization image
only 4\,ms after the original image, and image with a cycling
transition at an appropriate light level.}.

We image the atoms in the extremal $|f\!=\!9/2,m_f\!=\!-9/2\rangle$
state using a cycling transition. With a laser intensity of 12\% of
the saturation intensity $I_{sat}$, the atoms each scatter, on
average, 65 photons during our 40-$\mu$s pulse. We use a
back-illuminated CCD camera with a quantum efficiency of
$QE\!=\!80\%$. Additional losses on the imaging path give an
overall quantum efficiency of about 70\%. Figure 1(a) shows a raw
absorption image of an atom cloud.

To extract the noise signal from the absorption image,
$OD(\bm{r})$, we subtract an azimuthal average $\langle
OD(r)\rangle$ from each pixel to get a raw noise image $\Delta
OD(\bm{r})=OD(\bm{r})-\langle OD(r)\rangle$. In looking for atom
shot noise, the size of the image pixels plays an important role.
The atom number on a pixel or bin with area $a_{bin}^2$ is given by
$N_a\!=\!\langle OD(r)\rangle\,a_{bin}^2/\sigma$. For Poissonian
atom shot noise the magnitude of the fractional noise on a bin
given by $\delta OD_\mathrm{atom}/OD=\sqrt{N_a}/N_a$ increases for
smaller bins. However, if the bins are too small, the measured
noise will be reduced by the fact that the absorption signal due to
a single atom will be spread out over a finite area. In our
experiment causes of this blurring include the imaging resolution
($\approx 5\,\mu$m), the random walk motion of the atoms because of
photon scattering ($1\,\mu$m$-2\,\mu$m), and the falling motion of
the cloud during the imaging pulse (up to $7.5\,\mu$m). To study
the role of pixel size we can effectively vary the size of a bin
for a given image by averaging over multiple camera pixels. We do
this smoothly by applying a spatial low-pass filter with a variable
cutoff spatial frequency $f_\mathrm{lp}$. In addition, we apply a
high pass filter at a cutoff spatial frequency
$f_\mathrm{hp}=f_\mathrm{lp}/4$ to eliminate technical noise that
occurs at low spatial frequencies. For our filter (see
\footnote{After a 2D Fourier transform, a Butterworth Bandpass
filter function
$\left(1+\left(f/f_\mathrm{lp}\right)^{2n}\right)^{-1}-\left(1+\left(f/f_\mathrm{hp}\right)^{2n}\right)^{-1}$
is applied, where $n=5$ is the filter order and $f$ is the norm of
the spatial frequency. Then the image is transformed back to real
space, simultaneously increasing the image resolution by
interpolation.}), a numerical simulation yields an effective bin
size of $a_{bin}=1.25/f_\mathrm{lp}$.

Figure 1(b) shows a processed noise image, and Fig. 1(c)
demonstrates that atom shot noise can be the dominant noise source.
For medium bin sizes, the measured noise at the cloud center
($OD$$\approx$1) is close to the expected atom shot noise. For
small bin sizes, the noise is lower than expected. We attribute
this to the blurring described above. For very large bin sizes,
technical noise becomes more important, and the measured noise
exceeds the expected atom shot noise. The measured background noise
outside the atom cloud is only slightly larger than the expected
photon shot noise, which is given by $\delta
OD_\mathrm{photon}=\sqrt{1+e^{OD}}/\sqrt{N_p}$, where $N_p$ is the
photon number per effective bin.

The experiments probing pair-correlated atoms are initiated by
trapping and cooling a dilute gas of fermionic $^{40}$K atoms to
ultralow temperatures \cite{Regal2003b,DeMarco1999a}. We initially
prepare a nearly equal, incoherent mixture of atoms in the
$|f,m_f\rangle=|9/2,-9/2\rangle$ and $|9/2,-7/2\rangle$ spin
states, where $f$ is the total spin and $m_f$ the magnetic
sublevel. The atoms, and the molecules we create from these atoms,
are confined in a cigar-shaped far off-resonant optical dipole trap
with radial trapping frequencies on the order of 300\,Hz and an
axial frequency of $\nu_z\approx\nu_r/70$.

Weakly bound molecules are created by tuning the interaction
between atoms in the two spin states with a $7.8\pm0.6$\,G
\cite{Regal2003a} wide $s$-wave magnetic-field Feshbach resonance
at $202.10\pm0.07\,$G \cite{Regal2004a}. Slowly sweeping the
magnetic field across the Feshbach resonance results in a pairwise
conversion of $85\pm 5\%$ of the atoms into molecules
\cite{Hodby2004a}. For adiabatic $B$-field ramps and our lowest
temperatures, we get a Bose-Einstein condensate of molecules with
typically 15\% condensate fraction \cite{Regal2004a}.
Pair-correlated atoms are then produced by dissociating the
molecules with one of two techniques discussed below. The resulting
atoms form an entangled singlet pair, with one atom in each of the
two initial spin states. This entanglement follows from required
exchange symmetry of the fermionic atoms and the s-wave nature of
the interaction.

In a first experiment, we look for spatial atom-atom correlations
by probing the gas immediately after the dissociation of weakly
bound diatomic molecules. Here we ballistically expand a cloud of
about $3 \times10^5$ molecules for 19.2 ms. Then, we quickly
dissociate the molecules into two atoms in the $m_f\!=\!-7/2$ and
$m_f\!=\!-9/2$ states by increasing the magnetic field across the
Feshbach resonance. Immediately after the dissociation, we image
both spin states separately as described below. If the atoms do not
move significantly relative to each other \footnote{The atoms move
vertically because of gravity between the two images. We separately
determine the cloud center with a 2D surface fit in both
pictures.}, the atom shot noise for the $m_f\!=\!-7/2$ and
$m_f\!=\!-9/2$ images should be nearly identical.

To probe the singlet state, we need to independently measure atoms
in two spin states quasi-instantaneously. This is done with a
sequence of two pictures taken within 340\,$\mu$s using a kinetics
mode of the CCD camera. In the first absorption image only atoms in
the $m_f\!=\!-9/2$ state are addressed. Because of a large Zeeman
shift, absorption by atoms in other spin states is negligible. In
the second picture, we selectively probe atoms in the
$m_f\!=\!-7/2$ or $m_f\!=\!-5/2$ state by first flipping their
spins to the extremal $m_f\!=\!-9/2$ state with one or two
radio-frequency (rf) $\pi$-pulses, respectively, and then imaging
in the $m_f\!=\!-9/2$ state. The Rabi rates for these rf
transitions are, on average, $\Omega\!=\!2\pi\times30\,$kHz.

\begin{figure}[h]
\begin{center}
\includegraphics[width=1\linewidth]{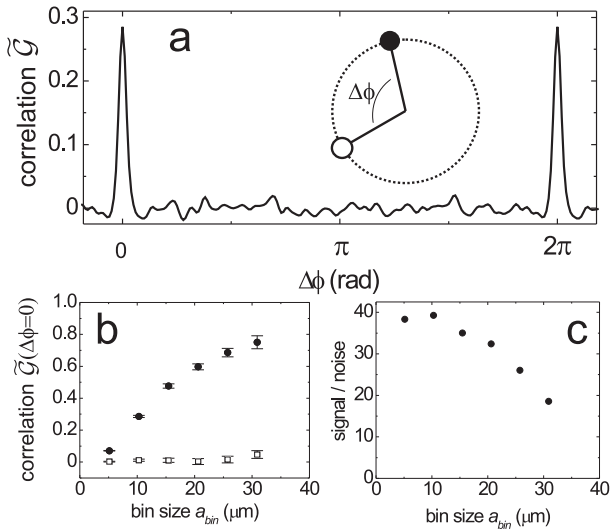}
\caption{Pair-correlated atoms. We plot the measured noise
correlation as a function of a relative angle of rotation between
absorption images of atoms in the two spin states (inset), averaged
over 11 images. The effective bin size is 10.3$\,\mu$m. Spatial
pair correlations $\tilde{\mathcal{G}}_{\alpha\beta}(0)$ can
clearly be seen when molecules are dissociated after expansion and
then the atoms are immediately imaged [(a) and $\bullet$ in (b)].
The correlation signal disappears when the molecules are
dissociated during an early stage of expansion [$\circ$ in (b)]. By
changing the low-pass filter to correspond to larger effective bin
sizes, we measure a correlation signal as large as $75\%$ (b).
However, the signal-to-noise ratio in detecting the correlation is
maximized for a lower effective bin size (c). }
\end{center}
\end{figure}

Figure 2 shows that we can clearly observe the atom-atom
correlations using the noise. We consider the correlation function
$\mathcal{G}_{\alpha\beta}(\bm{r},\bm{r}')\!=\!\Delta
N_\alpha(\bm{r})\Delta N_\beta(\bm{r}')$, where $\alpha$ and
$\beta$ denote the imaged spin states and $\Delta
N_\alpha(\bm{r})=\Delta OD_\alpha(\bm{r})a^2_{bin}/\sigma$ is the
fluctuation of atom number per bin, for atoms in spin state
$\alpha$. We calculate the correlation between atoms in the spin
states $\alpha$ and $\beta$ as a function of an angle representing
a relative rotation about the cloud center. Specifically, we write
the position in polar coordinates as $\bm r\!=\!(r,\phi)$ and $\bm
r'\!=\!(r',\phi+\Delta\phi)$, centered on the cloud, and calculate
the normalized correlation profile as
\begin{equation}
\tilde{\mathcal{G}}_{\alpha\beta}(\Delta\phi)=\left\langle\frac{\left\langle\,\Delta
N_\alpha(r,\phi)\,\,\Delta
N_\beta(r,\phi+\Delta\phi)\,\right\rangle_{\phi}}
{\sqrt{\overline{N}_\alpha(r) \overline{N}_\beta(r)} }
\right\rangle_{r}.
\end{equation}
Here, $\overline{N}_\alpha(r)$ is the azimuthally averaged number
of atoms in spin-state $\alpha$ per effective bin, and the
correlations are averaged over the angle $\phi$, normalized, and
then radially averaged. $\tilde{\mathcal{G}}_{\alpha\beta}(0)$ is
the correlation between spin states $\alpha$ and $\beta$ on
identical points in space, and
$\tilde{\mathcal{G}}_{\alpha\beta}(\pi)$ for diametrically opposite
points. For perfect correlations $\Delta
N_\alpha(r,\phi)\!=\!\Delta N_\beta(r,\phi+\Delta\phi)$ and
Poissonian atom shot noise $\delta
N_\alpha\!=\!\sqrt{\overline{N}_\alpha}$, the correlation function
$\tilde{\mathcal{G}}_{\alpha\beta}(\Delta\phi)$ is unity.

If the molecules are dissociated immediately before imaging, we
find a clear positive correlation signal for $\Delta\phi\!=\!0$
[Fig.\,2(a)]. In looking at the dependence of the correlation peak
on the effective bin size in Figs.\,2(b) and 2(c), we find that a
small bin size is optimal for detecting the presence of
correlations, while a larger bin size is necessary to accurately
measure the amount of correlation. We find the largest correlation
signal of 0.75 for our largest bin sizes \footnote{The correlation
signal has a relative systematic uncertainty of $\approx 25$\,\%
because of the uncertainty in measuring the absolute atom number.},
which is close to an expected signal of 0.85 for our $85\pm 5\%$
molecule conversion efficiency. The lower signal for smaller bins
indicates that the correlations are spread out over a finite area
in the images. One cause of this could be the relative motion of
atoms in the time between dissociation and imaging of the second
spin state. Indeed, we find no correlations if we dissociate the
atoms immediately after the start of TOF expansion [Fig. 2(b)]. On
the other hand, the signal-to-noise ratio for detecting the
presence of the pair correlations, defined as the ratio of
$\tilde{\mathcal{G}}_{\alpha\beta}(0)$ to the standard deviation of
$\tilde{\mathcal{G}}_{\alpha\beta}(\Delta\phi\neq 0)$, is maximized
for a significantly smaller effective bin size. This is because
smaller bins have larger fractional atom number noise and one has
more pixel pairs over which to average the correlation signal. The
width of the correlation peaks is limited by the effective bin size
and the blurring of the correlations.

\begin{figure}[h]
\begin{center}
\includegraphics[width=0.7\linewidth]{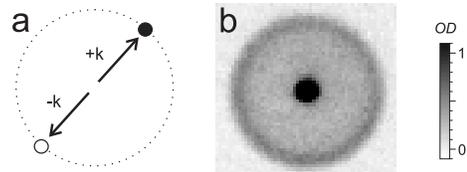}
\caption{(a) Atoms with equal but opposite momentum are found on
opposite sides of the atom cloud in TOF expansion. (b) This atom
absorption image was taken after rf photodissociation of weakly
bound molecules using an rf detuning of $\Delta \nu_{rf}\!=\!1.3$
MHz. The pair-correlated atoms comprise an expanding spherical
shell, containing approximately 1.3$\times10^5$ atoms per spin
state, which appears as a ring in the 2D absorption image.}
\end{center}
\end{figure}

In a second experiment, we are able to detect nonlocal pair
correlations between atoms that have equal but opposite momentum
and are therefore found at diametrically opposite points of the
atom cloud in TOF expansion [Fig.\,3(a)]. These pair correlations
are created by dissociation of molecules in the optical trap and
expansion of the atom gas before imaging. A significant further
experimental challenge arises from the fact that any center-of-mass
motion of the pairs rapidly degrades the correlation signal due to
blurring. We use several strategies to minimize this effect. First,
we start with an ultracold molecule sample. Second, although we
start with molecules in the strongly interacting regime at
$B\!=\!202.07$\,G, we rapidly change the B-field to 198\,G (within
$50\,\mu s$) before dissociation to avoid strong interaction
effects during the expansion. Third, we dissociate the molecules to
non-zero relative momentum states using rf photodissociation
\cite{Regal2003c}.

Detuning the rf with respect to the atomic transition by
$\Delta\nu_{rf}+E_{binding}$ results in free atoms (now in the
$m_f\!=\!-9/2$ and $m_f\!=\!-5/2$ spin-states) that fly apart in
opposite directions with a relative momentum of
$p\!=\!\sqrt{m\,h\,\Delta\nu_{rf}}$. By increasing the rf detuning,
we can give the atoms a relative momentum that is much larger then
their center-of-mass momentum [Fig. 3(b)]. In the TOF images, we
expect $\tilde{\mathcal{G}}_{(-5/2,-9/2)}(\Delta\phi)$ to show
positive correlation at $\Delta\phi\!=\!\pi$.

\begin{figure}[h]
\begin{center}
\includegraphics[width=1\linewidth]{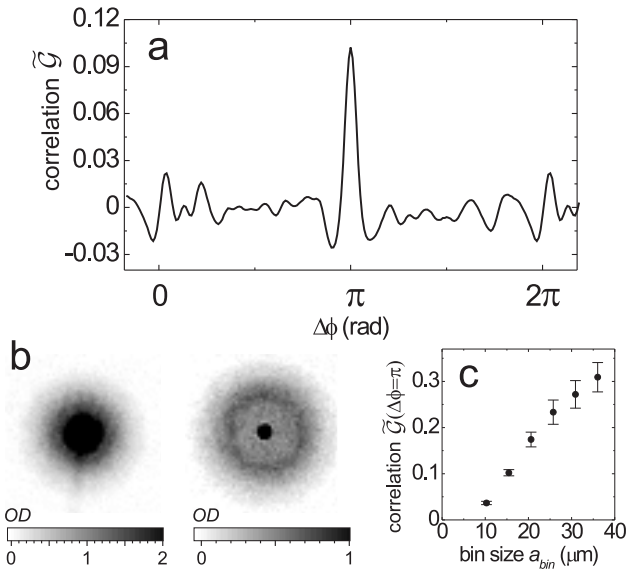}
\caption{Atom pair correlations in momentum space. (a) The averaged
correlation signal for 102 image pairs shows a peak for atoms with
equal but opposite momentum. The effective bin size is
15.5$\,\mu$m. (b) TOF absorption images of atoms in the two
spin-states, taken after 1.4\,ms and 1.7 ms. (c) For large
effective bins, we observe a correlation signal as high as 30\%.
The signal-to-noise ratio is on the order of 10.}
\end{center}
\end{figure}

Figure 4(b) shows example absorption images used to find nonlocal
correlations in the experiment. Here the molecules are dissociated
using a 330-$\mu s$ rf pulse with an average rf detuning of
$\Delta\nu_{rf}\!=\!1.1$\,MHz beyond the dissociation threshold. In
order to spread the dissociation shell over more camera pixels, we
sweep the rf over 600\,kHz. The absorption images are then taken
after only 1.4\,ms and 1.7\,ms of expansion. Before calculating the
correlation signal, we radially scale the second image to account
for the additional expansion time. Figure 4(a) shows a correlation
signal $\tilde{\mathcal{G}}_{(-5/2,-9/2)}(\Delta\phi)$ averaged for
102 images \footnote{To further optimize the signal-to-noise ratio,
we have subtracted off the technical correlations that are constant
over all images. These are determined by first averaging the 102
images and then calculating the remaining correlation signal.}. The
clear peak at $\Delta\phi=\pi$ corresponds to nonlocal correlations
in position, which are a result of correlations in momentum space.
We find that including a BEC in our ultracold molecular sample does
not significantly increase the correlation strength.

With a similar method \cite{Altman2004a}, it seems feasible to
directly probe generalized Cooper pairs in the BCS-BEC crossover
region \cite{Regal2004a,Zwierlein2004a,Chin2004a}. These pairs
would be detected as momentum correlations in the same way as
presented here. For this measurement, it will be important to
maximize the ratio between the relative and the center-of-mass
momentum of the dissociated pairs and minimize the collision rate
during the initial stage of TOF expansion. In the experiments we
have done thus far we find that the magnitude of the correlation
signal critically depends on these two parameters.

In conclusion, we have demonstrated that analysis of the noise in
absorption imaging can be used to directly probe atom-atom
correlations in a quantum gas. Here we have been able to detect
both local and nonlocal atom pair correlations created by the
dissociation of weakly bound singlet molecules. The new method
presented here will allow probing of interesting many-body states
such as Cooper pairs in an atomic Fermi gas as well as
anti-ferromagnetic phases and spin waves in optical lattices
\cite{Altman2004a}.

We would like to thank Ehud Altman for fruitful discussions and
Jason Smith for experimental assistance. This work was supported by
NSF and NASA. C. A. R. acknowledges support from the Hertz
Foundation.

\bibliographystyle{prsty}

\end{document}